\documentclass[
hf,
]{ceurart}

\usepackage{listings}
\usepackage{comment}
\usepackage{graphicx}
\usepackage{subcaption}

\begin{document}

\copyrightyear{2024}
\copyrightclause{Copyright for this paper by its authors.
  Use permitted under Creative Commons License Attribution 4.0
  International (CC BY 4.0).}

\title{Shorts on the Rise: Assessing the Effects of YouTube Shorts on Long-Form Video Content}

\author[1]{Prajit T. Rajendran}[%
email=prajit.rajendran@upfluence.com,
]

\address[1]{Upfluence,
Upfluence Inc.,
214 Sullivan Street,
Suite 3A,
New York City, NY 10012
USA}

\author[1]{Kevin Creusy}[%
email=kevin.creusy@upfluence.com,
]

\author[1]{Vivien Garnes}[%
email=vivien.garnes@upfluence.com,
]

\conference{}

\renewcommand{\shortauthors}{Prajit T Rajendran, et al.}

\begin{abstract}
Short form content has permeated into the video creator space over the past few years, led by industry leading products such as TikTok, YouTube Shorts and Instagram Reels. YouTube in particular was previously synonymous with being the main hub for long form video content consumption. The monetization of long form videos was easier as it allowed multiple advertisement placements during the course of the video. This model also facilitated thematic brand partnerships. However, since the introduction of short form content, creators have found it more difficult to generate revenue as advertisement placements have decreased. This leads to a unique situation where people are spending more time watching shorter videos, and yet they generate less revenue for the creators. In this paper, we perform a study of 250 creators with significant audiences to see if the introduction of short form content has affected the view counts and engagement of long form content. Our findings reveal a noteworthy trend: since the advent of short-form content, there has been a significant decrease in both view counts and engagement in long-form videos on these channels.
\end{abstract}


\maketitle

\section{Introduction}

In the ever-evolving landscape of digital media, the surge of short-form content has marked a significant shift in the video creation and consumption space. Historically, digital platforms like YouTube have been the main hub for long-form video content, offering creators lucrative financial benefits through multiple advertisement placements and thematic brand partnerships. This model not only propped up creators financially but also fostered a rich environment for narrative depth and audience engagement. However, the recent advent of short-form content platforms such as TikTok, YouTube Shorts, and Instagram Reels, has disrupted this established norm.\\

The introduction of short-form content has been both rapid and pervasive, reshaping the expectations and consumption habits of audiences globally. These platforms, characterized by their bite-sized content cater to the decreasing attention spans of the modern digital audience. The shift has been significant not just in content length, but also in the nature of monetization. The traditional long-form content allowed for multiple advertisement insertions, leading to more significant revenue generation for creators. In contrast, short-form content, with its limited duration, offers fewer opportunities for advertisement placements, posing a challenge for creators accustomed to the long-form revenue model.\\

It is reported by TechCrunch \cite{tech_crunch}, an American global online newspaper, that as of 2023 more than 2 billion logged-in monthly users were watching YouTube Shorts. This is up from 1 billion in late 2021, giving YouTube Shorts an edge over competitors like TikTok and Instagram Reels. Based on industry averages estimated by Wyzowl \cite{YT_stats}, a creator needs around 500,000 views to earn \$1,000 from advertisements, and the cost per thousand (CPM) is expected be \$2 per 1,000 impressions.\\

The YouTube Partner Program (YPP) has been central in transforming YouTube into a platform where professionally generated content more aligned with the interests of advertisers takes precedence as elaborated by R. Lobato et.al \cite{lobato2016cultural}. As per J. Kim et.al \cite{kim2012institutionalization}, it is estimated that YouTube shares 55\% of the revenue generated by ads shown on monetized videos with their creators. Advertisement revenue is accessible to creators as long as they adhere to the guidelines for “advertiser-friendly content” that prohibit the use of inappropriate language, depiction of violence, drug use, and other things not considered suitable for advertisers \cite{YT_answer}. The YPP has recently evolved to include different advertisement formats and now offers different monetization features inside the platform such as channel memberships through which viewers support individual channels directly\cite{rieder2023making}. The YPP also accepts product placements, sponsorships and endorsements in their videos provided that the creators include appropriate disclosure. Creators are also known to include affiliate links in their video descriptions, which acts as an additional source of income. Long form content brought with it a plethora of options for monetization- be it through an increased viewing of advertisements by the audience, product placements or sponsorships. However, with the advent of short form content which makes product placement and sponsored content difficult for the most part due to time restrictions, monetization has become increasingly difficult for creators.\\

This new era presents a fresh set of challenges for creators. On one hand, there is an undeniable increase in the consumption of short-form videos; they are quick, engaging, and align well with the fast-paced lifestyle of the contemporary audience. On the other hand, this shift raises critical questions about the financial sustainability for creators, especially those who have built their careers around long-form content. The ease of creating short-form content and its widespread appeal has led to a saturated market, where standing out and monetizing effectively becomes increasingly challenging.\\

\subsection{Contributions:}
This paper aims to unpack these findings in detail, exploring the implications of short-form content for content creators and the industry at large. We analyze the changing dynamics of viewer behavior, the adaptation strategies employed by creators, and the broader economic implications of this shift. Through this study, we aim to provide valuable insights into the evolving nature of digital content creation and its future trajectory. The contributions of this paper are as follows:
\begin{itemize}
    \item Analysis of video statistics of popular creators to study the effect of short form content on the engagement in long form content
    \item Analysis of video statistics of popular creators to study the differences in consumption of long form content and short form content
    \item Analysis of video statistics of popular creators to study the differences in consumption between different cohorts of creators
\end{itemize}

\section{Related works}
The influence of short-form content on long-form content consumption and creation is a subject of growing interest within the digital media research community. Various studies have explored different facets of this evolving landscape, from viewer engagement patterns to monetization strategies and the psychological impact of content length on audience preferences. This section reviews relevant literature to contextualize our research within the broader field of digital media studies.\\

DV Dunas et.al (2020) \cite{dunas2020emerging} and Alzubi (2023) \cite{alzubi2023evolving} revealed that media consumption amongst the younger generation had a clear leaning toward digital media, especially when it came to social media, streaming services, and online news websites and digital media. The accessibility and enormous content library prompt the selection of digital media as the default choice for both entertainment and information. It is clear that the proliferation of mobile devices has significantly contributed to the popularity of short-form content within digital media, due to its convenience and accessibility. This sets the stage for understanding the technological and social catalysts behind the rise of platforms like TikTok and Youtube Shorts. \\

AC Munaro et.al \cite{munaro2021engage} note that long-form content tends to generate deeper viewer involvement, as evidenced by more meaningful comments and higher viewer retention rates. It is also shown that medium-length and long videos posted during non-business hours and weekdays and those using a subjective language style, less-active events, and temporal indications are more likely to receive high engagement. S. Wu et.al \cite{wu2018beyond} illustrate that video duration is the primary source of variation for engagement in videos. This explains why short-form content typically enjoys higher number of views than long form content which requires a dedication of more time from the viewer. \\

B Rieder et. al \cite{rieder2023making} provide an in-depth analysis of monetization strategies employed by creators on YouTube. This work contrasts the traditional ad-based revenue model of long-form content with the emerging trends in short-form content, including sponsored posts, brand partnerships, and platform-specific monetization features. For instance, the study explains that 16-17\% of YouTube videos employ crowdfunding or marketplace links as revenue sources. These results highlight the challenges creators face in adapting their revenue strategies to accommodate the constraints of short-form content. \\

W Maenhout \cite{maenhoutlength} performed an experimental study that examines the effectiveness of advertisements concerning long-form versus short-form video advertisements. This study found that there is no significant impact in the relationship between video length (long form vs.
short form) and advertising effectiveness. This contradicts the hypothesis that a long-form video advertisement is more conducive to create a more positive attitude towards the brand
than a short-form advertisement, further illustrating why platforms push for short form content. Their findings underscore the pressure on long-form creators to adapt their content strategies, echoing the concerns about decreased view counts and engagement highlighted in our research.\\

As per SC Kies \cite{kies2018social}, frequent exposure to short form content was shown to have an effect on sustained attention span in the younger generation, particularly in relation to distractibility. MZ Newman et.al \cite{newman2010new}, TJ Kohler \cite{kohler2023caught} also discuss that short-form videos can affect attention span and mood in viewers. The addictive nature of short form content thus contributes to their popularity amongst younger audiences, and is one of the primary reasons for the shift of platforms from long form content to short form content.\\

Our research builds upon these foundational studies, providing empirical evidence on the specific impacts of short-form content introduction on the engagement and viewership of long-form videos among creators with significant audiences. By focusing on a quantitative analysis of engagement metrics for videos of selected creators, our study aims to contribute a nuanced understanding of how short-form content is reshaping the landscape of digital video content.

\section{Methodology}
This study aims to elucidate the impact of short-form content on the engagement and viewership of long-form videos among YouTube creators. To achieve this objective, we employed a mixed-methods approach, combining quantitative data extraction with comparative analysis. Our methodology encompasses several steps, from data collection using the data from YouTube and the Upfluence database of influencers to detailed analytical comparisons of video engagement metrics. This section outlines the methodology employed in our research, detailing the data collection process, the selection criteria for YouTube creators, and the analytical methods used to assess the impact of YouTube Shorts on long-form content.

\subsection{Data Collection}
\begin{figure}[tbh]
  \centering
  \includegraphics[width=0.6\linewidth]{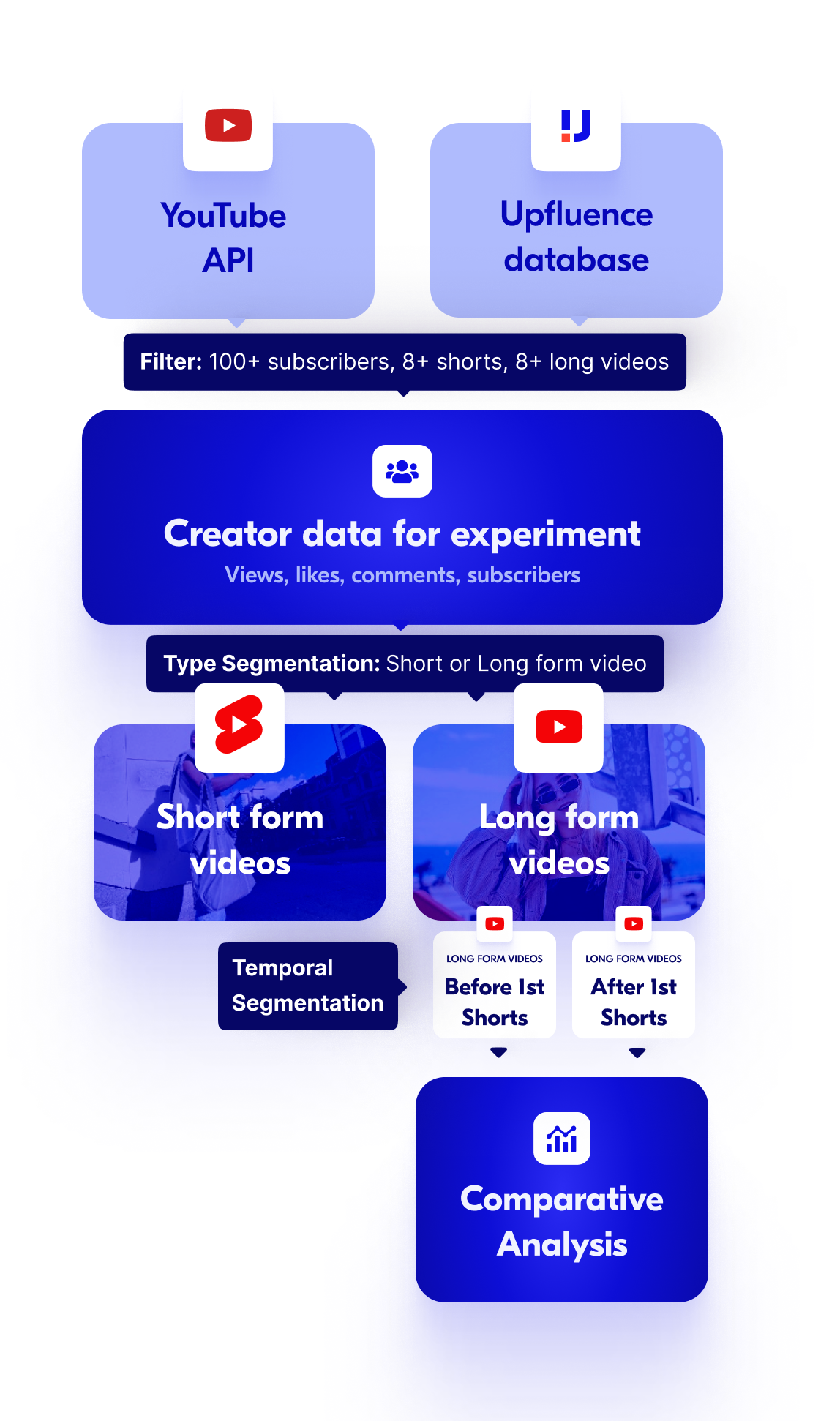}
  \caption{Block diagram depicting the data collection and analysis followed in this study}
  \label{fig:blk_diag}
\end{figure}

We collected data from 250 prominent creators across various fields, such as entertainment, education, lifestyle, and tech. The selection criteria for these creators were based on their subscriber count, with a focus on those considered to have significant audience reach (defined here as having 100,000 subscribers or more). This criterion ensured the inclusion of creators with substantial impact and visibility on the platform. Another constraint was to only include creators who have published at least 8 short form videos and 8 long form videos. The analysis was restricted to videos released before December 31, 2023. Care was taken to include creators from different categories, and with different subscriber counts. Additionally, we identified the first publication date of a YouTube Shorts video for each creator to serve as a temporal marker for our comparative analysis. This date served as a pivotal point for our analysis, enabling a before-and-after comparison of long-form content engagement. The identification of this date allowed us to examine the immediate and long-term effects of integrating short-form content into a creator's video portfolio. Figure \ref{fig:blk_diag} illustrates the process of data collection and other steps followed in this study.\\

The data extracted for each creator included:
\begin{itemize}
    \item A complete list of published videos, categorized into short-form (YouTube Shorts) and long-form content.
    \item Engagement metrics for each video, specifically views, likes, and comments.
    \item The publication date of each video, which was critical for identifying the temporal relationship between the introduction of short-form content and changes in long-form video engagement.
\end{itemize}

\begin{figure}
  \centering
  \includegraphics[width=0.7\linewidth]{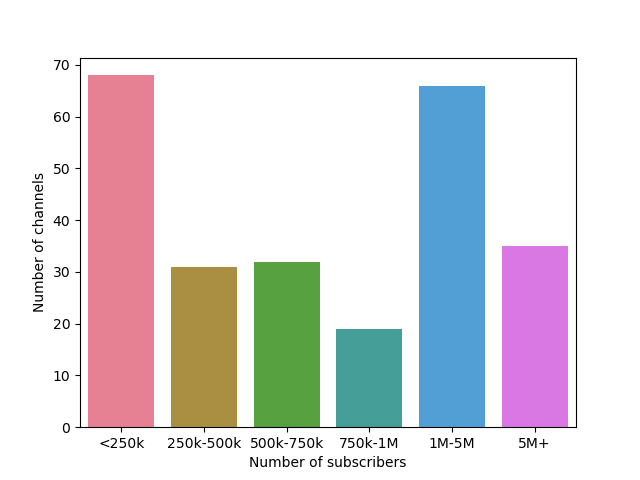}
  \caption{Number of channels across different subscriber categories}
  \label{fig:subscribers}
\end{figure}

\begin{figure}
  \centering
  \includegraphics[width=0.7\linewidth]{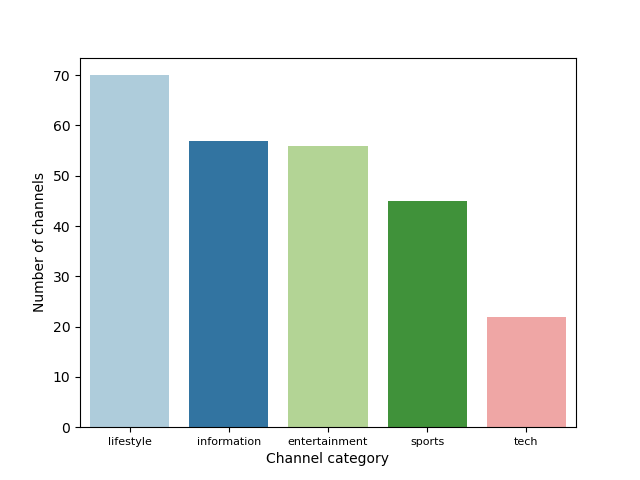}
  \caption{Number of channels across different channel categories}
  \label{fig:channels}
\end{figure}

Figure \ref{fig:subscribers} illustrates the distribution of subscriber counts amongst the selected channels. We divided the creators into six categories: those with <250k subscribers, 250k-500k subscribers, 500k-750k subscribers, 750k-1 million subscribers, 1 million-5 million subscribers, and those with more than 5 million subscribers. Figure \ref{fig:channels} depicts the distribution of channel categories amongst the selected channels. This includes categories such as lifestyle, entertainment, sports, technology and information. \\

\subsection{Data pre-processing}
The raw data obtained underwent a pre-processing phase to ensure accuracy and relevance for the analysis. This process involved:

\begin{itemize}
    \item \textbf{Data Cleaning:} Removal of incomplete or anomalous data entries that could skew the results.
    \item \textbf{Categorization of Content:} Videos were classified into two categories: short-form (YouTube Shorts) and long-form content, based on their duration. YouTube Shorts were defined as videos lasting up to 60 seconds, aligning with YouTube's criteria for short-form content.
    \item \textbf{Temporal Segmentation:} For each creator, videos were segmented into two temporal groups based on the publication date of their first YouTube Shorts video. This segmentation resulted in two datasets for long-form content: one comprising videos published before the first Shorts video on the channel and another comprising those published afterward. 
\end{itemize}

\subsection{Research focus}
Our analysis consisted of three primary comparative evaluations:

\begin{itemize}
    \item \textbf{Short-form vs. Long-form Content Engagement:} We calculated the engagement metrics such as view counts for both short-form (YouTube Shorts) and long-form videos across all creators. This comparison aimed to assess the overall engagement difference between short-form and long-form content on the platform.
    \item \textbf{Impact of YouTube Shorts on Long-form Content:} We performed a before-and-after comparison for long-form videos, analyzing the engagement metrics for videos released before and after the publication of the first YouTube Shorts video by each creator. This analysis sought to identify any shifts in engagement metrics for long-form content following the creators' adoption of YouTube Shorts.
    \item \textbf{Impact of creator category on how YouTube Shorts are adopted:} We performed a comparison amongst different creator categories to study the adoption of YouTube Shorts on their channel. This analysis sought to identify if certain categories of creators were affected more by the introduction of short form content.
\end{itemize}

To determine the significance of the observed differences in engagement metrics, we employed the Wilcoxon signed-rank statistical test to compare the mean values of engagement between the defined groups. The level of significance was set at p < 0.05. \\

\section{Analysis and findings}
As elaborated in section 3.1, data from 250 creators was collected and analyzed in this work. Considering the absolute number of views for a video does not account for the temporal advantage of older videos to have been present on the internet for longer to be discovered as opposed to more recently videos. To account for this bias, we normalize the counts of views with an exponential decay function as follows:
\begin{equation}
    v_{adjusted} = v_{raw} / (days^\alpha)
\end{equation}

where $v_{raw}$ is the raw view count of the video, $days$ is the number of days since the video was published and $\alpha$ is the decay factor. A larger value of $\alpha$ overly compensates older videos whereas not using a decay function fails to consider that the older videos have a temporal advantage of having been present longer. The value of $\alpha$ was arrived at 0.05 experimentally in this study. \\

\begin{figure}[tbh]
\centering
    \begin{subfigure}[b]{0.6\textwidth}
        \includegraphics[width=\linewidth]{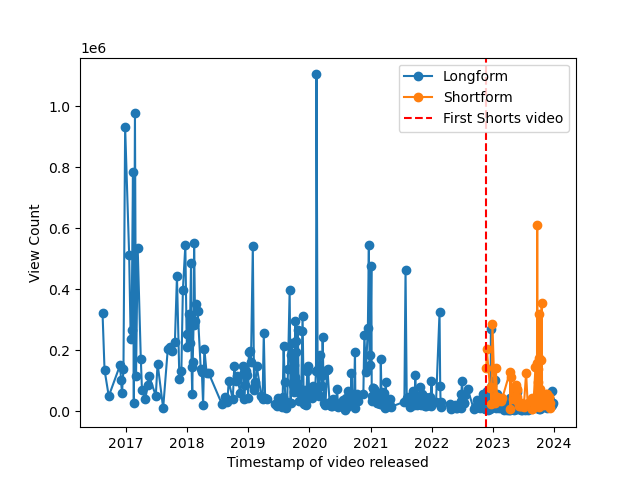}
        \caption{}
        \label{fig:sub1}
    \end{subfigure}

    \begin{subfigure}[c]{0.6\textwidth}
        \includegraphics[width=\linewidth]{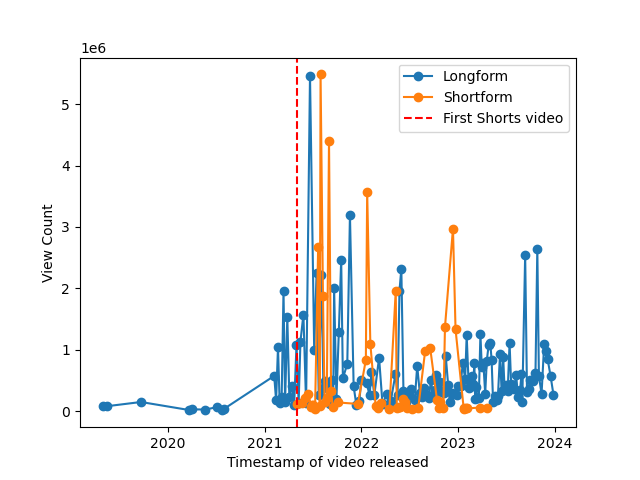}
        \caption{}
        \label{fig:sub2}
    \end{subfigure}
    \caption{View count on channels (a)'HockeyStigg' and (b)'HowMoneyWorks' plotted over time}
    \label{fig:positive}
\end{figure}

Figure \ref{fig:positive} shows two different scenarios that were observed amongst the creators. Some creators as depicted in (a) started focusing more on short form content and their long form content dropped both in terms of views and number of videos released. However, other creators such as (b) did not pursue short form content or did not achieve much success in it, leading to their long form views not being affected as much.

\subsection{Analysis of long form content on the whole data}
Looking at the data as a whole, it was observed that there was a mean decrease per channel in long form content views by 743,589 since the first Shorts video published on those respective channels. The data distribution was not a normal distribution, as verified by the popular Shapiro-Wilk Normality test. A Wilcoxon signed-rank test was performed to test the statistical significance of the results. It was observed that the p-value was 0.011, which is lower than the significance level set at 0.05, indicating that \textbf{there is a significant difference in long form views before and after the introduction of Shorts}. In terms of other engagement metrics, our study did not elucidate with statistical significance that the number of likes and comments decreased since the introduction of Shorts.\\

\noindent Summary: On the data overall, we can see that the there was a decrease in the long form content views after the introduction of Shorts on the channel.

\subsection{Analysis on creator types- majority short form or long form}
Out of the 250 creators, 71 were creators who produce a larger number of Shorts videos than long form videos. As the study has an objective to analyze the performance of long form content, we divide the creators into two cohorts: ones who produce more short form content (S) and ones who produce more long form content (L). These two cohorts had different statistics as described in table \ref{table:stats_cohort}. In this study, creators in the cohort S produced 420 Shorts and 128 long form videos on average as opposed to creators in cohort L who produced 103 Shorts and 632 long form videos on average. The analysis here includes 38,297 and 132,393 videos extracted from the creators of cohorts S and L respectively. \\

\begin{table}
  \caption{Statistics of the S and L cohorts}
  \label{tab:commands}
  \begin{tabular}{ccc c l}
    \toprule
    Cohort & \# of creators & Avg. short video count & Avg. long video count & Total \# of videos\\
    \midrule
    S & 71 & 420 &128 & 38,927\\
    L & 179 & 103 &632 & 132,393\\
    \bottomrule
  \end{tabular}
  \label{table:stats_cohort}
\end{table}

Analysing the cohort 'L', it can be seen that there was a mean decrease per channel in long form content views by 920,189 since the first Shorts video published on those respective channels. The Wilcoxon signed-rank test resulted in a p-value of 0.005, which is lower than the significance level set at 0.05, indicating that \textbf{there is a significant difference in long form views before and after the introduction of Shorts}. However, with the cohort 'S', the results are different. The mean decrease per channel was observed to be 163,813 and with a p-value>0.05, it \textbf{does not indicate that there was a significant decrease in the number of views for this cohort}. This is expected as creators who create mostly long form content are the ones who had to shift their strategy to start making short form content. This is a sharp contrast to creators who focused on making short form content from the start, as they did not particularly have many views for their long form content in the first place.\\

\begin{figure}
  \centering
  \includegraphics[width=0.7\linewidth]{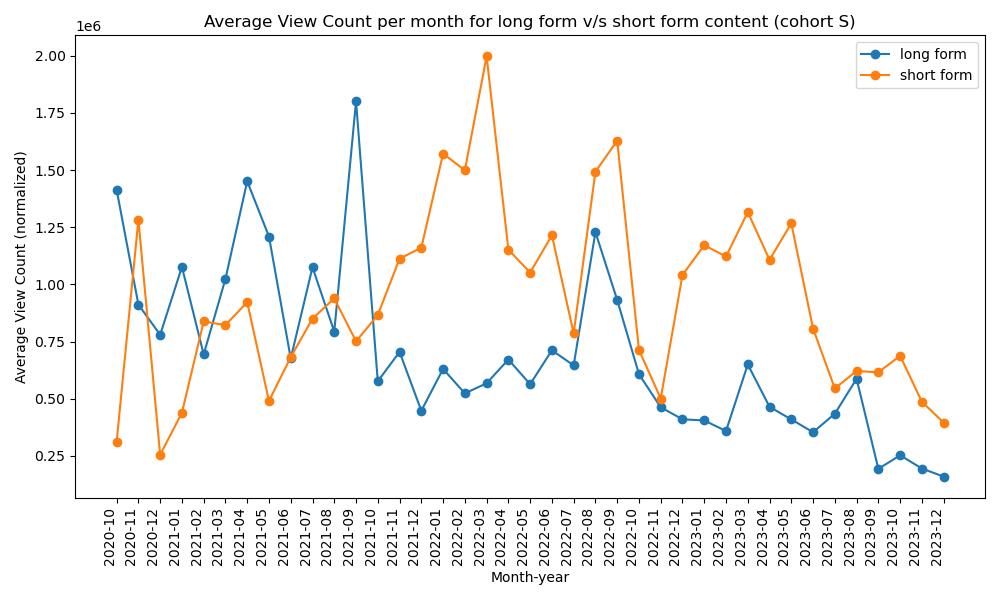}
  \caption{Average view count since the introduction of Shorts on YouTube (cohort L)}
  \label{fig:long_major}
\end{figure}

\begin{figure}
  \centering
  \includegraphics[width=0.7\linewidth]{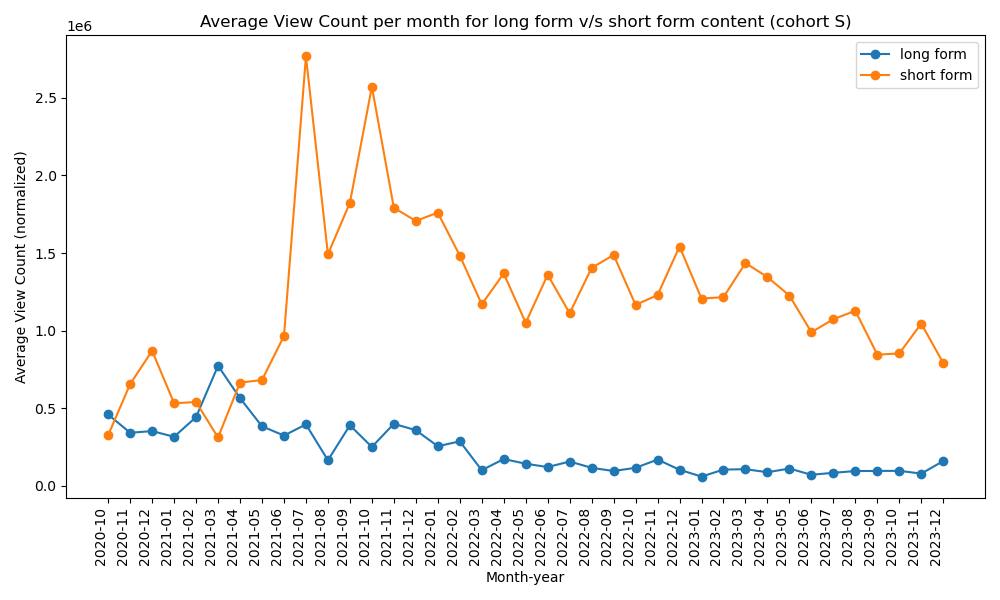}
  \caption{Average view count since the introduction of Shorts on YouTube(cohort S)}
  \label{fig:short_major}
\end{figure}

Figures \ref{fig:long_major} and \ref{fig:short_major} illustrates the distribution in views for long form and short form videos since the introduction of YouTube Shorts in the fourth quarter of 2020 for cohorts L and S respectively. In the context of creators who mostly create long form content, it can be seen that short form content produced by them is now performing better than their long form content, which gives them incentive to switch over to short form content. When we look at creators who mostly make short form content, it can be seen that they are justified in their decision to make more short form videos since they perform much better than the long form content that they produce. Hence, it makes sense for creators in both cohorts to produce short form content in order to obtain higher viewership counts. \\

\noindent Summary: For creators who make mostly long form content, we can see that the there was a decrease in the long form content views after the introduction of Shorts on the channel. On the contrary for creators who make mostly short form content, we can observe that there was no statistically significant decrease in long form views after the introduction of Shorts on the channel. 

\subsection{Analysis on creator types- subscriber categories}
\begin{figure}
  \centering
  \includegraphics[width=0.7\linewidth]{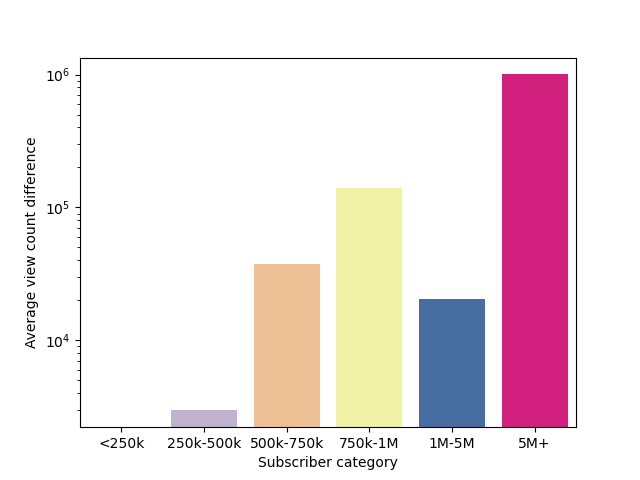}
  \caption{Average view count difference across different subscribers categories}
  \label{fig:view_subscribers}
\end{figure}

Table \ref{table:stats_cohort} describes the statistics of creators divided into different subscriber count categories. Figure \ref{fig:view_subscribers} shows the average decrease in views after the introduction of Shorts per subscriber count category. It was observed that for smaller creators, the decrease in views was not significant, due to their audience size and typical number of views being lower. For creators with >5 million subscribers, we found the result to be statistically significant using the Wilcoxon signed-rank test. For the other subscriber categories, the results were not statistically significant. \\

\noindent Summary: For very larger creator with over 5 million subscribers, the effect of decrease in views in long form content after the increase in short form content was more profound. This has significant implications in terms of advertisement as the top creators account for a majority of views overall on YouTube. 

\begin{table}
  \caption{Statistics of the subscriber category cohorts}
  \label{tab:commands}
  \begin{tabular}{ccc c l}
    \toprule
    Subscriber count & \# of creators & Avg. short video count & Avg. long video count & Total \# of videos\\
    \midrule
    <250k & 68 & 158 &387 & 37,123\\
    250k-500k & 31 & 229 &375 & 18,735\\
    500k-750k & 32 & 211 &360 & 18,310\\
    750k-1M & 19 & 243 &456 & 13,307\\
    1M-5M & 65 & 207 &690 & 58,332\\
    >5M & 35 & 163 &565 & 25,513\\
    \bottomrule
  \end{tabular}
  \label{table:stats_cohort}
\end{table}

\begin{figure}
  \centering
  \includegraphics[width=0.7\linewidth]{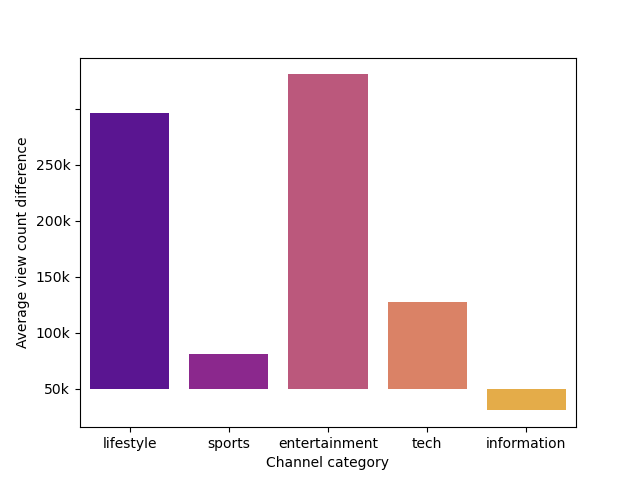}
  \caption{Average view count difference across different channel categories}
  \label{fig:view_channels}
\end{figure}

\subsection{Analysis on creator types- channel categories}
YouTube has a listed set of channel categories based on the content of the channel, but since the list is broad for our study, we combined some of the categories to form super-categories. The following are the categories we selected in this work:
\begin{itemize}
    \item \textbf{Lifestyle:} Included channels with the YouTube assigned categories of 'How-to and style',     'Pets and animals' and 'Travel and events'
    \item \textbf{Information:} Included channels with the YouTube assigned categories of 'Education', 'News and politics'
    \item \textbf{Entertainment:} Included channels with the YouTube assigned categories of 'Comedy', 'Entertainment', 'Film and animation' and 'Music'
    \item \textbf{Sports:} Included channels with the YouTube assigned categories of 'Sports'
    \item \textbf{Technology:} Included channels with the YouTube assigned categories of 'Autos and vehicles' and 'Science and Technology'
\end{itemize}
Table \ref{table:stats_cohort} describes the statistics of creators divided into different channel categories. Figure \ref{fig:view_channels} depicts the average decrease in views after the introduction of Shorts in each channel category. It can be seen that the long form view counts for lifestyle and entertainment related content decreased drastically as opposed to the categories tech, sports or information. As informational or educational content is mostly long form, Shorts have not eaten into the share of views in this category as much as it has in categories like entertainment and lifestyle, which is more conducive to short form content. \\

\noindent Summary: The introduction of Shorts has resulted in a decrease in long form view counts in certain categories such as entertainment and lifestyle, while not significantly affecting educational channels. 

\begin{table}
  \caption{Statistics of the channel category cohorts}
  \label{tab:commands}
  \begin{tabular}{ccc c l}
    \toprule
    Channel category & \# of creators & Avg. short video count & Avg. long video count & Total \# of videos\\
    \midrule
    Lifestyle & 70 &209 &370 & 40,635\\
    Information & 57 & 151 &394 & 31,082\\
    Entertainment & 56 & 234 &364 & 33,532\\
    Sports & 45 & 190 &825 & 45,751\\
    Technology & 22 & 156 &767 & 20,320\\
    \bottomrule 
  \end{tabular}
  \label{table:stats_cohort}
\end{table}

\subsection{Analysis of short form and long form content}
Looking at the data as a whole, table \ref{table:overall_views} shows that short form content performs better on average than long form content in terms of views. It can be seen that long form creators have had to adapt to the changing environment and produce more short form content whereas creators who primarily operated in short form content are generating plenty of views and are not delving into long form content. As for likes, table \ref{table:overall_likes} shows that short form content enjoys more number likes as opposed to long form content on average. This may be due to the quick nature of consumption of Shorts viewed in portrait mode on devices, where the viewer can quickly tap the like button present prominently on the bottom right part of the screen and move on to the next video. In contrast, long form content is mostly viewed in landscape mode on mobile devices and the like button is not as prominent as it is for short form content. However, when it comes to comments, the trend is reversed. Table \ref{table:overall_comments} describes that long form content enjoys more number likes as opposed to short form content on average. This is again perhaps due to the quick nature of consumption of short form content where viewers swipe and move on to the next video whereas long form content allows for more time to write comments. \\

\noindent Summary: Short form content performs better than long form content in terms of views and likes on average. However, long form content sees more comments as opposed to short form content. 

\begin{table}
  \caption{Comparing views for long form content and short form content over various categories}
  \label{tab:commands}
  \begin{tabular}{ccc c l}
    \toprule
    Category & Avg. long form views  & Avg. short form views\\
    \midrule
    Overall & 755,128 & 983,394\\
    Long form creators (L) & 950,439 & 914,549\\
    Short form creators (S) & 246,447 & 1,151,399\\
    \bottomrule
  \end{tabular}
  \label{table:overall_views}
\end{table}

\begin{table}
  \caption{Comparing likes for long form content and short form content over various categories}
  \label{tab:commands}
  \begin{tabular}{ccc c l}
    \toprule
    Category & Avg. long form likes  & Avg. short form likes\\
    \midrule
    Overall & 15,855 & 52,252\\
    Long form creators (L) & 18,851 & 44,828\\
    Short form creators (S) & 8,427 & 70,550\\
    \bottomrule
  \end{tabular}
  \label{table:overall_likes}
\end{table}

\begin{table}
  \caption{Comparing comments for long form content and short form content over various categories}
  \label{tab:commands}
  \begin{tabular}{ccc c l}
    \toprule
    Category & Avg. long form comments  & Avg. short form comments\\
    \midrule
    Overall & 1,030 & 402\\
    Long form creators (L) & 1303 & 403\\
    Short form creators (S) & 353 & 398\\
    \bottomrule
  \end{tabular}
  \label{table:overall_comments}
\end{table}

\section{Conclusions}
The results of our study support the hypothesis that short form content generates more engagement than long form content. The results also demonstrate that long form content released since the advent of short form content on a specific channel have suffered a reduction in engagement as opposed to those released on the channel prior to the introduction of short form content. A cohort analysis among different groups of creators also illustrated that there is a difference in the pattern of adoption of short form content between different categories of channels. These findings have implications for the creator community as they support the claims of creators that novel monetization mechanisms need to be created to account for the changes in video consumption patterns on platforms such as YouTube.\\ 

\bibliography{sample-ceur}

\begin{thebibliography}{14}
\expandafter\ifx\csname natexlab\endcsname\relax\def\natexlab#1{#1}\fi
\providecommand{\url}[1]{\texttt{#1}}
\providecommand{\href}[2]{#2}
\providecommand{\path}[1]{#1}
\providecommand{\DOIprefix}{doi:}
\providecommand{\ArXivprefix}{arXiv:}
\providecommand{\URLprefix}{URL: }
\providecommand{\Pubmedprefix}{pmid:}
\providecommand{\doi}[1]{\href{http://dx.doi.org/#1}{\path{#1}}}
\providecommand{\Pubmed}[1]{\href{pmid:#1}{\path{#1}}}
\providecommand{\bibinfo}[2]{#2}
\ifx\xfnm\relax \def\xfnm[#1]{\unskip,\space#1}\fi
\bibitem[{Mehta(2023)}]{tech_crunch}
\bibinfo{author}{I.~Mehta}, \bibinfo{title}{Google says 2 billion logged in monthly users are watching youtube shorts}, \bibinfo{year}{2023}. \URLprefix \url{https://techcrunch.com/2023/07/25/google-says-2-billion-logged-in-monthly-users-are-watching-youtube-shorts/}.
\bibitem[{Hayes(2024)}]{YT_stats}
\bibinfo{author}{A.~Hayes}, \bibinfo{title}{Youtube stats: Everything you need to know in 2024!}, \bibinfo{year}{2024}. \URLprefix \url{https://www.wyzowl.com/youtube-stats/}.
\bibitem[{Lobato(2016)}]{lobato2016cultural}
\bibinfo{author}{R.~Lobato},
\newblock \bibinfo{title}{The cultural logic of digital intermediaries: Youtube multichannel networks},
\newblock \bibinfo{journal}{Convergence} \bibinfo{volume}{22} (\bibinfo{year}{2016}) \bibinfo{pages}{348--360}.
\bibitem[{Kim(2012)}]{kim2012institutionalization}
\bibinfo{author}{J.~Kim},
\newblock \bibinfo{title}{The institutionalization of youtube: From user-generated content to professionally generated content},
\newblock \bibinfo{journal}{Media, culture \& society} \bibinfo{volume}{34} (\bibinfo{year}{2012}) \bibinfo{pages}{53--67}.
\bibitem[{YouTube(2024)}]{YT_answer}
\bibinfo{author}{YouTube}, \bibinfo{title}{Advertiser-friendly content guidelines}, \bibinfo{year}{2024}. \URLprefix \url{https://support.google.com/youtube/answer/6162278}.
\bibitem[{Rieder et~al.(2023)Rieder, Borra, Coromina, and Matamoros-Fern{\'a}ndez}]{rieder2023making}
\bibinfo{author}{B.~Rieder}, \bibinfo{author}{E.~Borra}, \bibinfo{author}{{\`O}.~Coromina}, \bibinfo{author}{A.~Matamoros-Fern{\'a}ndez},
\newblock \bibinfo{title}{Making a living in the creator economy: A large-scale study of linking on youtube},
\newblock \bibinfo{journal}{Social Media+ Society} \bibinfo{volume}{9} (\bibinfo{year}{2023}) \bibinfo{pages}{20563051231180628}.
\bibitem[{Dunas and Vartanov(2020)}]{dunas2020emerging}
\bibinfo{author}{D.~V. Dunas}, \bibinfo{author}{S.~A. Vartanov},
\newblock \bibinfo{title}{Emerging digital media culture in russia: modeling the media consumption of generation z},
\newblock \bibinfo{journal}{Journal of Multicultural Discourses} \bibinfo{volume}{15} (\bibinfo{year}{2020}) \bibinfo{pages}{186--203}.
\bibitem[{Alzubi(2023)}]{alzubi2023evolving}
\bibinfo{author}{A.~Alzubi},
\newblock \bibinfo{title}{The evolving relationship between digital and conventional media: A study of media consumption habits in the digital era},
\newblock \bibinfo{journal}{THE PROGRESS: A Journal of Multidisciplinary Studies} \bibinfo{volume}{4} (\bibinfo{year}{2023}) \bibinfo{pages}{1--13}.
\bibitem[{Munaro et~al.(2021)Munaro, H{\"u}bner~Barcelos, Francisco~Maffezzolli, Santos~Rodrigues, and Cabrera~Paraiso}]{munaro2021engage}
\bibinfo{author}{A.~C. Munaro}, \bibinfo{author}{R.~H{\"u}bner~Barcelos}, \bibinfo{author}{E.~C. Francisco~Maffezzolli}, \bibinfo{author}{J.~P. Santos~Rodrigues}, \bibinfo{author}{E.~Cabrera~Paraiso},
\newblock \bibinfo{title}{To engage or not engage? the features of video content on youtube affecting digital consumer engagement},
\newblock \bibinfo{journal}{Journal of consumer behaviour} \bibinfo{volume}{20} (\bibinfo{year}{2021}) \bibinfo{pages}{1336--1352}.
\bibitem[{Wu et~al.(2018)Wu, Rizoiu, and Xie}]{wu2018beyond}
\bibinfo{author}{S.~Wu}, \bibinfo{author}{M.-A. Rizoiu}, \bibinfo{author}{L.~Xie},
\newblock \bibinfo{title}{Beyond views: Measuring and predicting engagement in online videos},
\newblock in: \bibinfo{booktitle}{Proceedings of the International AAAI Conference on Web and Social Media}, volume~\bibinfo{volume}{12}, \bibinfo{year}{2018}.
\bibitem[{Maenhout(????)}]{maenhoutlength}
\bibinfo{author}{W.~Maenhout},
\newblock \bibinfo{title}{Length does matter: The effectiveness of long-form versus short-form video marketing according to advertising context and viewer age}  (????).
\bibitem[{Kies et~al.(2018)}]{kies2018social}
\bibinfo{author}{S.~C. Kies}, et~al.,
\newblock \bibinfo{title}{Social media impact on attention span},
\newblock \bibinfo{journal}{Journal of Management \& Engineering Integration} \bibinfo{volume}{11} (\bibinfo{year}{2018}) \bibinfo{pages}{20--27}.
\bibitem[{Newman(2010)}]{newman2010new}
\bibinfo{author}{M.~Z. Newman},
\newblock \bibinfo{title}{New media, young audiences and discourses of attention: from sesame street to ‘snack culture’},
\newblock \bibinfo{journal}{Media, Culture \& Society} \bibinfo{volume}{32} (\bibinfo{year}{2010}) \bibinfo{pages}{581--596}.
\bibitem[{Kohler(2023)}]{kohler2023caught}
\bibinfo{author}{T.~J. Kohler}, \bibinfo{title}{Caught In The Loop: The Effects of The Addictive Nature Of Short-form Videos On Users’ Perceived Attention Span And Mood}, \bibinfo{type}{{B.S.} thesis}, University of Twente, \bibinfo{year}{2023}.

\end{thebibliography}

\end{document}